\documentclass[manuscript,screen,nonacm]{acmart} 
\setcopyright{cc} 
\acmConference[]{} 

\usepackage[utf8]{inputenc}
\usepackage{listings}
\usepackage{array,graphicx}
\usepackage{float}
\usepackage{multirow}

\newcommand{\workshopname}{GenAICHI: CHI 2025 Workshop on Generative AI and HCI}
\newcommand\extrafootertext[1]{
    \bgroup
    \renewcommand\thefootnote{\fnsymbol{footnote}}%
    \renewcommand\thempfootnote{\fnsymbol{mpfootnote}}%
    \footnotetext[0]{#1}%
    \egroup
}

\AtBeginDocument{ 
    \fancypagestyle{firstpagestyle}{
        \fancyhf{}
        \fancyfoot[L]{\sffamily\footnotesize \workshopname}%
        \fancyfoot[C]{\sffamily\footnotesize \thepage}
    }
    \fancyhf{}
    \fancyhead[L]{\sffamily\footnotesize\shorttitle}
    \fancyhead[R]{\sffamily\footnotesize\shortauthors}
    \fancyfoot[L]{\sffamily\footnotesize\workshopname}%
    \fancyfoot[C]{\sffamily\footnotesize\thepage}
}

\begin{document}

\title{A Case Study of Scalable Content Annotation Using Multi-LLM Consensus and Human Review}

\author{Mingyue Yuan}
\email{mingyue.yuan@unsw.edu.au}
\affiliation{%
\institution{ University of New South Wale}
 \country{Australia}
}

\author{Jieshan Chen}
\email{Jieshan.Chen@data61.csiro.au}
\affiliation{%
  \institution{CSIRO’s Data61 }
  \country{Australia}
}

 \author{Zhenchang Xing}
 \email{zhenchang.xing@data61.csiro.au}
\affiliation{%
  \institution{CSIRO’s Data61}
  \country{Australia}
}

\author{Gelareh Mohammadi}
\email{g.mohammadi@unsw.edu.au}
\affiliation{%
  \institution{University of New South Wales}
  \country{Australia}
}

\author{Aaron Quigley}
\email{a.quigley@data61.csiro.au}
\affiliation{%
  \institution{CSIRO’s Data61 }
  \country{Australia}
}

\begin{abstract}

Content annotation at scale remains challenging, requiring substantial human expertise and effort. This paper presents a case study in code documentation analysis, where we explore the balance between automation efficiency and annotation accuracy.
We present MCHR (Multi-LLM Consensus with Human Review), a novel semi-automated framework that enhances annotation scalability through the systematic integration of multiple LLMs and targeted human review. Our framework introduces a structured consensus-building mechanism among LLMs and an adaptive review protocol that strategically engages human expertise.
Through our case study, we demonstrate that MCHR reduces annotation time by 32\% to 100\% compared to manual annotation while maintaining high accuracy (85.5\% to 98\%) across different difficulty levels, from basic binary classification to challenging open-set scenarios.

\end{abstract}

 

\maketitle

\section{Introduction}

Generative AI, particularly Large Language Models (LLMs), has demonstrated impressive capabilities in text generation\cite{iqbal2022survey}, code completion\cite{jimenez2023swe}, and question answering\cite{weidinger2022taxonomy}, with growing applications in automating content analysis and annotation tasks \cite{park2024collaborative}\cite{pham2023topicgpt}. 
Traditional manual annotation methods are labor-intensive and time-consuming, especially for large-scale datasets like software repositories or multi-modal content\cite{liu2024using}\cite{gebreegziabher2023patat}. 
While automated AI annotation offers scalability, it struggles with accuracy and reliability \cite{zhang2024rethinking}\cite{lu2024we}. Therefore, exploring human-AI collaborative approaches to identify more effective collaborative paradigms is crucial.

This study presents a semi-automated annotation framework that leverages multiple LLMs for collaborative content analysis while maintaining human oversight. To explore new paradigms in human-AI collaboration, our framework addresses three critical challenges in content annotation:

First, while existing tools like quallm\cite{crone2024quallm} demonstrate basic LLM-assisted annotation capabilities, our framework advances this concept by introducing a structured multiple-LLM approach that systematically addresses classification tasks of varying difficulty levels. Previous work\cite{liu2024using}\cite{pham2023topicgpt} primarily focused on identifying outliers in specific domains. In contrast, we innovate by developing and evaluating a systematic framework that handles four distinct difficulty levels, ranging from basic binary classification to open-set categorization. Our evaluation demonstrates clear performance patterns across these levels - achieving over 98\% accuracy in basic tasks, 95.5\% in domain classification, 94.06\% in closed-set classification, and 85.5\% in challenging open-set scenarios. This evaluation reveals both the capabilities and limitations of LLM-assisted annotation across different complexity levels, providing practical insights for implementing semi-automated annotation tasks.

Second, we extend beyond traditional human-AI collaboration methods by implementing a novel consensus-building mechanism among multiple LLMs. While previous work in incivility annotation \cite{park2024collaborative} and medical diagnostics\cite{zhang2024rethinking} demonstrated basic benefits of combining AI and human judgment, our framework introduces a systematic approach to reconciling model disagreements. Unlike single-model systems\cite{crone2024quallm}\cite{gebreegziabher2023patat}, our approach leverages diverse LLM perspectives to provide diverse initial analyses, enabling even non-expert annotators to make informed decisions. This multi-model consensus approach significantly outperforms single-model baselines, particularly in complex tasks - improving accuracy by 8-32 percentage points for closed-set classification and doubling the accuracy (85.5\% vs 45.2\%) for open-set classification compared to the best single model. These results demonstrate the effectiveness of leveraging diverse model perspectives for enhanced annotation quality.

Third, we provide concrete evidence for automation reliability and workload reduction through analysis of human review requirements. Our framework achieves high automation reliability (91.5-99.1\% accuracy) for closed-set tasks (Levels 1-3) without human intervention, while strategic human review improves accuracy to 96\% for complex open-set tasks. Compared to full manual annotation, we reduce human workload by 100\%, 92\%, 66\%, and 32\% for Level 1-4 tasks respectively, while maintaining high quality standards. These results demonstrate a practical balance between automation efficiency and annotation accuracy, with human expertise strategically deployed where it provides the most value.


\section{Background}
\label{sec:Background}
\begin{figure*}
    \centering
    \includegraphics[width=1\linewidth]{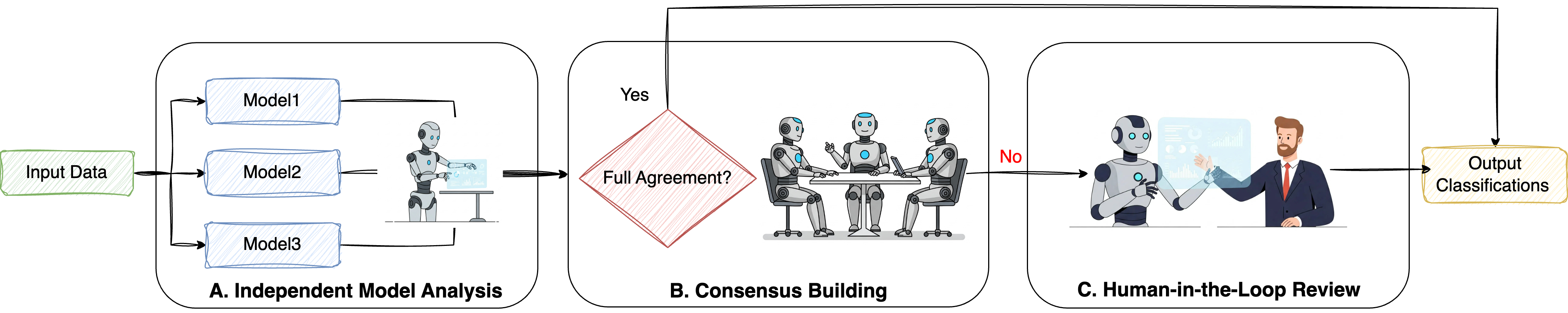}
    \caption{Overview of our semi-automated annotation framework: (A) Independent Model Analysis - multiple LLMs process input data independently; (B) Consensus Building - models collaborate to reach agreement; (C) Human-in-the-Loop Review - expert review for cases requiring human judgment.}
    \label{fig:overview}
\end{figure*}

In modern software development environments, teams face the challenge of rapidly categorizing\cite{farr2024llm} and routing various technical content, such as feature requests, and code changes and bug reports\cite{ahmed2021capbug}. For example, in enterprise settings, support staff need to quickly classify incoming technical issues across diverse domains - from web development to hardware concerns. Web development alone typically encompasses multiple specialized areas: Frontend development, Backend systems, Full-stack applications, Database management and Supporting tools and infrastructure.

Beyond these predefined categories, issues may relate to specialized domains like hardware interfaces, configuration management, or domain-specific applications. The challenge lies in efficiently categorizing this content while handling both common and specialized cases.

Our case study builds upon the COMMITPACK dataset introduced by Muennighoff et al.\cite{muennighoff2023octopack}, which contains 4TB of Git commits across 350 programming languages. This dataset pairs code changes with human-written commit messages, originally designed to improve instruction-based fine-tuning of LLMs. While the dataset was initially created for tasks like code fixing, interpretation, and synthesis, we identified its potential for exploring automated content classification in software development contexts. We use the high-quality subset COMMITPACKFT, which includes 277 programming languages. For each language, we randomly selected 10 files for our case study analysis, ensuring a diverse yet manageable dataset for thorough evaluation.

Our study examines three increasingly complex classification tasks using this dataset. The first task focuses on \textbf{Level 1 - basic classification} through binary language identification. This involves rapid assessment of programming languages (such as distinguishing JavaScript from non-JavaScript code) to enable initial triage and routing of technical issues. This fundamental step is crucial for directing issues to appropriate technical teams and establishing the basic context for further analysis.

The second task addresses \textbf{Level 2 - domain classification}, which involves categorizing code into broader application domains such as web development. This high-level categorization helps in department-level issue assignment and considers cross-domain implications. Such classification is essential for resource allocation and team coordination, particularly in large organizations where multiple teams may need to collaborate on complex issues.

The third and most complex task involves {both detailed categorization within identified \textbf{Level 3 - closed-set classification and Level 4 - open-set classification for emerging categories}. Within established domains like web development, this includes precise classification into areas such as frontend development, backend systems, full-stack applications, database management, and infrastructure support. However, equally important is the system's ability to identify and categorize content that falls outside these predefined categories. This open-set classification capability is crucial for handling the "long tail" of specialized technical issues, such as domain-specific frameworks, emerging technologies, or hybrid approaches that combine multiple paradigms. The system must not only recognize when content doesn't fit existing categories but also propose meaningful new categories and maintain consistency in how these novel cases are classified. The focus on both structured categorization and open-set recognition enables the system to evolve alongside rapidly changing technical landscapes while maintaining organizational efficiency.

Through these tasks, we aim to understand not only the technical capabilities of LLM-assisted content classification but also its practical utility in real-world software development environments. This case study allows us to evaluate both the technical feasibility and operational benefits of integrating LLM assistance into content classification workflows.

\section{Semi-Automated Annotation Framework}

We present a novel Multi-LLM Consensus with Human Review (MCHR) framework that leverages multiple Large Language Models (LLMs) for content analysis while maintaining human oversight. As shown in Figure \ref{fig:overview}, our framework consists of three main modules: independent model analysis, consensus building, and human-in-the-loop review.

\noindent
\textbf{Independent Model Analysis: }
Our framework employs three state-of-the-art LLMs (GPT-4o\cite{gpt4o}, Claude 3.5 Sonnet\cite{claude35}, and GPT-o1 reasoning\cite{gpto1_reasoning}) to analyze content independently, ensuring unbiased and diverse perspectives. Each model processes the same input using identical prompts. 
This independent processing approach helps capture diverse interpretations and reduces systematic biases that might arise from sequential or dependent analysis.

\noindent
\textbf{Consensus Building: }
Drawing inspiration from collaborative human annotation practices, we developed a systematic consensus-building mechanism that simulates collective expert decision-making through multiple LLMs. The protocol implements a three-stage verification process where two primary models initially analyze the content independently, with a third model serving as an additional evaluator in cases of initial disagreement. The system categorizes consensus outcomes into three levels: full agreement where all models provide the same result, partial agreement where two models agree while one differs, and no agreement where all models provide different classifications. This granular approach to consensus measurement enables informed decisions about the necessity of human intervention, particularly for cases of complete divergence or potential novel categories in open-set classification scenarios.




\noindent
\textbf{Human-in-the-Loop Review:}
The human review module is strategically activated in specific annotation scenarios: model disagreement or confidence scores lower than a threshold (0.8) in cases of partial agreement.
To facilitate informed decision-making, the system presents annotators with a comprehensive LLM analysis content, including model classifications, reasoning chains, and divergence points.
To ensure consistent annotation standards while minimizing unnecessary human intervention, we implemented several quality control modules. This system includes structured response validation to ensure output format consistency, confidence score thresholding for classifications, and regular quality checks through random sampling of agreed classifications.

Our framework tries to strike an optimal balance between automation efficiency and annotation accuracy across all four classification difficulty levels. Through configurable analysis prompts and adaptive classification schemes, the system maintains high performance across diverse content types. As illustrated in Figure \ref{fig:overview}, this systematic workflow facilitates seamless transitions between automated analysis, consensus building, and targeted human review, ensuring both scalability and reliability in the annotation process.

\section{Experimental Study}

In this section, we evaluate our Multi-LLM Consensus with Human Review framework's effectiveness for classification tasks, using three research questions (RQs):

\begin{itemize}
    \item \textbf{RQ1 (Performance Across Difficulty Levels):} How does the method perform across classification tasks of varying difficulty levels?

    \item \textbf{RQ2 (Multi-LLM Consensus Effectiveness):} How does the annotation accuracy of the Multi-LLM Consensus with Human Review framework compare to that of single-model baselines?

    \item \textbf{RQ3 (Automation Reliability):} Is the automated annotation part accurate enough ($\geq$90\%) to justify reducing human workload, and what is the impact of human review on accuracy in cases of model disagreement?

\end{itemize}

\subsection{Experimental Design}

As described in Section~\ref{sec:Background}, we conduct a case study analysis using the COMMITPACKFT dataset, a curated collection of code snippets representing 277 distinct programming languages. We randomly selected 10 files per language, yielding a total of 2,770 code snippets for use in our classification task analysis.

\noindent
\textbf{Difficulty Levels:} To answer \textbf{RQ1}, we defined a four-level classification for each code snippet. This level (corresponding to the classification tasks described in Section~\ref{sec:Background} allows us to systematically evaluate the system's capabilities across a wide spectrum of annotation challenges.

\textbf{Level 1 (Basic Classification):} Binary classification of code snippets (determining whether a snippet is \textit{JavaScript}).

\textbf{Level 2 (Domain Classification):} 
Determine the code's broad application domain (e.g., \textit{web development}) via binary classification.
 
\textbf{Level 3 (Closed-Set Classification):} Categorizing code snippets into one of five predefined categories: \textit{frontend, backend, full-stack, database, and supporting tools}.

\textbf{Level 4 (Open-Set Classification):} Identifying and categorizing code snippets that fall outside the predefined categories in Level 3. This level addresses the ``long tai'' of less common programming languages and tasks by proposing appropriate new category labels.

To establish a gold standard for evaluating the classification tasks, three software developers with over six years of industry experience independently annotated the dataset we selected. Two developers performed the initial annotation, using online resources and AI-assisted tools as needed. A third developer resolved conflicts by reviewing annotations in cases of disagreement. On average, each annotation took approximately three minutes.

\noindent
\textbf{Evaluation Setup:} 
We evaluated our Multi-LLM Consensus with Human Review (MCHR) framework against strong single-model baselines to assess its effectiveness.  The baselines consisted of the individual performances of state-of-the-art LLMs: GPT-4o, Claude 3.5 Sonnet, and GPT-o1 reasoning. For the MCHR framework itself, we used these same three models as Model 1, Model 2, and Model 3, respectively, to generate the multi-model consensus.

To address \textbf{RQ2}, we compared the overall performance of MCHR (denoted as \textit{``MCHR (All)''} in Table \ref{tab:level-performance}) against the single-model baselines.  This comparison demonstrates the overall benefit of our approach.

To address \textbf{RQ3}, we performed a more granular analysis, separately evaluating the MCHR (All) results into two parts:

\textbf{MCHR (Auto-part):} 
This represents the accuracy achieved solely by the automated multi-model in cases of full agreement, without any human review. It reflects the annotation performance of the system's automated part.

\textbf{MCHR (Human-part):}This presents the accuracy of only the cases that required human Review (due to model disagreement or low confidence). This isolate the accuracy of the human Review part.
By comparing MCHR (Auto-part) and MCHR (Human-part), and the MCHR (All), we can quantify the accuracy improvement provided by human intervention in disputed cases.

Finally, to quantify the reduction in human workload, we measured the Human Review Rate (HRR). This metric represents the percentage of cases that required human review because the models either disagreed or expressed low confidence in their predictions. A lower HRR indicates greater automation and reduced need for human intervention.

\begin{table}[ht]
\caption{Classification Performance Across Difficulty Levels (\% ± 95\% CI).  MCHR = Multi-LLM Consensus with Human Review.}
\label{tab:level-performance}
\centering
\small 
\setlength{\tabcolsep}{4pt} 
\begin{tabular}{@{}lcccc@{}}
\toprule
\textbf{Approach} & Level 1 & Level 2 & Level 3 & Level 4 \\
\midrule
GPT-4o\cite{gpt4o} & 98.1 ±1.8 & 92.5 ±3.2 & 62.1 ±11.4 & 29.9 ±6.3 \\
Claude 3.5 Sonnet\cite{claude35}  &  98.2 ±1.7 & 91.7 ±3.3 & 80.0 ±9.3 & 37.7 ±6.5 \\
GPT-o1\cite{gpto1_reasoning}   & 98.6 ±1.5 &  92.9 ±1.4 & 85.7 ±5.2 &45.2 ±5.0 \\
\midrule
MCHR (All)              & \textbf{98.1 ±1.8} & \textbf{95.5 ±2.6}    & \textbf{94.06 ±8.6}    & \textbf{85.5 ±5.1}    \\
MCHR (Auto-part) & 99.1 ± 1.2    & 95.2 ± 2.7    & 91.5 ± 8.5    & 83.6 ± 11.7    \\
MCHR (Human-part)            & - & 96.7 ±3.0     & 100.0 ±6.2     & 96.0 ±3.7 \\
\midrule
Human Review Rate  & 0.00          & 6.74          & 33.33         & 67.20         \\
\bottomrule
\end{tabular}
\end{table}

\subsection{Results and Analysis}

\textbf{RQ1 (Performance Across Difficulty Levels):} Our framework demonstrated severe performance degradation patterns across difficulty levels (Table~\ref{tab:level-performance}). For basic classification (Level 1), all approaches achieved near-perfect accuracy (98.1-98.6\%), showing LLMs' high performance in low-level classification. At Level 2 (domain classification), MCHR (All) achieved 95.5\% accuracy (±2.6), outperforming single models by 3-4 percentage points. The performance gap widened significantly at higher difficulty levels: for closed-set classification (Level 3), MCHR improved accuracy by 8-32 percentage points over baselines, and for open-set classification (Level 4), it doubled the accuracy of the best single model (85.5\% vs 45.2\%).

\noindent
\textbf{RQ2 (Multi-LLM Consensus Effectiveness):} The multi-LLM consensus mechanism substantially improved annotation quality, particularly for complex tasks. While single models showed high variance in Level 3 (62.1-85.7\%) and Level 4 (29.9-45.2\%) performance, MCHR (All) achieved stable high accuracy (94.06\% and 85.5\% respectively) through consensus building. 
The increasing human review rate with task complexity (from 33.33\% at Level 3 to 67.2\% at Level 4) underscores the need for this consensus approach.

\noindent
\textbf{RQ3 (Automation Reliability):} The MCHR Auto-part demonstrated high accuracy on Level 1-3 tasks (91.5-99.1\%), indicating reliable performance for closed-set problems and a reduction in manual annotation workload. However, the accuracy for open-set problems was 83.6\%, falling short of the 90\% acceptance criterion for high-quality automatic annotation. Manual review was crucial for improving accuracy in those challenging cases, resulting in an accuracy of 96\% for Level 4 tasks in MCHR (Human-part). 

Future research should investigate mechanisms to enhance both the human review rate and the accuracy of the automatic annotation part for such complex classification problems.

Compared to full manual annotation, the framework reduced manual workload by 100\%, 92\%, 66\%, and 32\% for Level 1, 2, 3, and 4 tasks, respectively, while maintaining a final accuracy above 85\% for both closed-set and challenging open-set annotation tasks through strategic human-AI collaboration.

\subsection{Key Findings and Future Directions}

\noindent
\textbf{Key Takeaway Findings:}
This study revealed several significant insights into human-AI collaborative content analysis. First, the multi-LLM consensus mechanism demonstrated remarkable effectiveness in mitigating individual model biases, particularly in ambiguous cases where single models showed a 1–23\% accuracy variance across different levels of task difficulty. This finding underscores the value of diverse model perspectives in improving annotation reliability.

Second, analysis of the 201 open-set cases revealed critical challenges in handling sparse categories. With 73\% of categories containing fewer than three samples (mean: 2.01), significant naming inconsistencies were observed across results. For example, semantically similar concepts received varied labels (e.g., ``hardware description'' versus ``HDL programming''), highlighting the need for robust taxonomy management in open-set classification.

Our approach proved highly efficient, reducing annotation time by 66\%–100\% compared to pure manual annotation while maintaining quality standards for closed-set classification analysis; furthermore, human verification served as both immediate quality control and a valuable alignment signal for the future research.

Our approach demonstrated a substantial reduction in annotation time (66\% to 100\%) relative to manual annotation, while preserving quality standards for closed-set classification. 
Moreover, human verification not only provided immediate quality control but also offered a valuable signal for future research improvement.


\noindent
\textbf{Future Potential Improvement:}
Building on our findings, we identify two critical areas for enhancement. First, the consensus mechanism could be strengthened through confidence-based model debates and dynamic weighting of model votes based on historical performance in the future. This direction would help address the current limitations in handling low-confidence predictions and improve the overall reliability of automated classifications.

Second, we envision a future adaptive learning pipeline that leverages human-verified examples through a few-shot learning method. This pipeline would create continuous feedback loops between human reviewers and models, gradually improving classification accuracy while reducing the need for human intervention. Such a potential approach could reinforce specialized domains where standard classification schemes may not fully capture the nuances of the content.

\noindent
\textbf{Future Applications:}
The framework's potential extends beyond our current implementation in programming language classification. In research community, it could significantly enhance systematic literature reviews and research data annotation, where consistent categorization across large volumes of content is crucial. The system's ability to handle both predefined categories and emerging concepts makes it particularly valuable for fields with rapidly evolving terminology and concepts.

In industrial aspect, the framework could transform technical support systems through intelligent ticket routing and bug report triage. Its ability to maintain classification consistency while adapting to new categories addresses a critical need in company environments where support issues often span multiple technical domains. The reduced annotation time and maintained accuracy levels make it particularly suitable for high-volume, time-sensitive classification tasks.
The human-in-the-loop module ensures that domain expertise guides the evolution of classification schemes while maximizing automation benefits.



\bibliographystyle{ACM-Reference-Format}
\bibliography{reference}


\begin{thebibliography}{16}


\ifx \showCODEN    \undefined \def \showCODEN     #1{\unskip}     \fi
\ifx \showDOI      \undefined \def \showDOI       #1{#1}\fi
\ifx \showISBNx    \undefined \def \showISBNx     #1{\unskip}     \fi
\ifx \showISBNxiii \undefined \def \showISBNxiii  #1{\unskip}     \fi
\ifx \showISSN     \undefined \def \showISSN      #1{\unskip}     \fi
\ifx \showLCCN     \undefined \def \showLCCN      #1{\unskip}     \fi
\ifx \shownote     \undefined \def \shownote      #1{#1}          \fi
\ifx \showarticletitle \undefined \def \showarticletitle #1{#1}   \fi
\ifx \showURL      \undefined \def \showURL       {\relax}        \fi
\providecommand\bibfield[2]{#2}
\providecommand\bibinfo[2]{#2}
\providecommand\natexlab[1]{#1}
\providecommand\showeprint[2][]{arXiv:#2}

\bibitem[Ahmed et~al\mbox{.}(2021)]%
        {ahmed2021capbug}
\bibfield{author}{\bibinfo{person}{Hafiza~Anisa Ahmed}, \bibinfo{person}{Narmeen~Zakaria Bawany}, {and} \bibinfo{person}{Jawwad~Ahmed Shamsi}.} \bibinfo{year}{2021}\natexlab{}.
\newblock \showarticletitle{Capbug-a framework for automatic bug categorization and prioritization using nlp and machine learning algorithms}.
\newblock \bibinfo{journal}{\emph{IEEE Access}}  \bibinfo{volume}{9} (\bibinfo{year}{2021}), \bibinfo{pages}{50496--50512}.
\newblock


\bibitem[Anthropic(2024)]%
        {claude35}
\bibfield{author}{\bibinfo{person}{Anthropic}.} \bibinfo{year}{2024}\natexlab{}.
\newblock \bibinfo{title}{Claude 3.5 Sonnet: Enhanced reasoning, state-of-the-art coding skills, computer use, and 200K context window}.
\newblock \bibinfo{howpublished}{\url{https://www.anthropic.com/claude/sonnet}}.
\newblock
\newblock
\shownote{Accessed: 2025-02-17}.


\bibitem[Crone(2024)]%
        {crone2024quallm}
\bibfield{author}{\bibinfo{person}{Damien Crone}.} \bibinfo{year}{2024}\natexlab{}.
\newblock \bibinfo{title}{quallm: A python library for LLM-assisted content coding}.
\newblock \bibinfo{howpublished}{\url{https://github.com/damiencrone/quallm}}.
\newblock
\newblock
\shownote{Accessed: 2025-02-17}.


\bibitem[Farr et~al\mbox{.}(2024)]%
        {farr2024llm}
\bibfield{author}{\bibinfo{person}{David Farr}, \bibinfo{person}{Nico Manzonelli}, \bibinfo{person}{Iain Cruickshank}, \bibinfo{person}{Kate Starbird}, {and} \bibinfo{person}{Jevin West}.} \bibinfo{year}{2024}\natexlab{}.
\newblock \showarticletitle{Llm chain ensembles for scalable and accurate data annotation}. In \bibinfo{booktitle}{\emph{2024 IEEE International Conference on Big Data (BigData)}}. IEEE, \bibinfo{pages}{2110--2118}.
\newblock


\bibitem[Gebreegziabher et~al\mbox{.}(2023)]%
        {gebreegziabher2023patat}
\bibfield{author}{\bibinfo{person}{Simret~Araya Gebreegziabher}, \bibinfo{person}{Zheng Zhang}, \bibinfo{person}{Xiaohang Tang}, \bibinfo{person}{Yihao Meng}, \bibinfo{person}{Elena~L Glassman}, {and} \bibinfo{person}{Toby Jia-Jun Li}.} \bibinfo{year}{2023}\natexlab{}.
\newblock \showarticletitle{Patat: Human-ai collaborative qualitative coding with explainable interactive rule synthesis}. In \bibinfo{booktitle}{\emph{Proceedings of the 2023 CHI Conference on Human Factors in Computing Systems}}. \bibinfo{pages}{1--19}.
\newblock


\bibitem[Iqbal and Qureshi(2022)]%
        {iqbal2022survey}
\bibfield{author}{\bibinfo{person}{Touseef Iqbal} {and} \bibinfo{person}{Shaima Qureshi}.} \bibinfo{year}{2022}\natexlab{}.
\newblock \showarticletitle{The survey: Text generation models in deep learning}.
\newblock \bibinfo{journal}{\emph{Journal of King Saud University-Computer and Information Sciences}} \bibinfo{volume}{34}, \bibinfo{number}{6} (\bibinfo{year}{2022}), \bibinfo{pages}{2515--2528}.
\newblock


\bibitem[Jimenez et~al\mbox{.}(2023)]%
        {jimenez2023swe}
\bibfield{author}{\bibinfo{person}{Carlos~E Jimenez}, \bibinfo{person}{John Yang}, \bibinfo{person}{Alexander Wettig}, \bibinfo{person}{Shunyu Yao}, \bibinfo{person}{Kexin Pei}, \bibinfo{person}{Ofir Press}, {and} \bibinfo{person}{Karthik Narasimhan}.} \bibinfo{year}{2023}\natexlab{}.
\newblock \showarticletitle{Swe-bench: Can language models resolve real-world github issues?}
\newblock \bibinfo{journal}{\emph{arXiv preprint arXiv:2310.06770}} (\bibinfo{year}{2023}).
\newblock


\bibitem[Liu et~al\mbox{.}(2024)]%
        {liu2024using}
\bibfield{author}{\bibinfo{person}{Jiaying Liu}, \bibinfo{person}{Yunlong Wang}, \bibinfo{person}{Yao Lyu}, \bibinfo{person}{Yiheng Su}, \bibinfo{person}{Shuo Niu}, \bibinfo{person}{Yan Zhang}, {et~al\mbox{.}}} \bibinfo{year}{2024}\natexlab{}.
\newblock \showarticletitle{Using Large Language Models to Assist Video Content Analysis: An Exploratory Study of Short Videos on Depression}.
\newblock \bibinfo{journal}{\emph{arXiv e-prints}} (\bibinfo{year}{2024}), \bibinfo{pages}{arXiv--2406}.
\newblock


\bibitem[Lu et~al\mbox{.}(2024)]%
        {lu2024we}
\bibfield{author}{\bibinfo{person}{Jinwei Lu}, \bibinfo{person}{Yikuan Yan}, \bibinfo{person}{Keman Huang}, \bibinfo{person}{Ming Yin}, {and} \bibinfo{person}{Fang Zhang}.} \bibinfo{year}{2024}\natexlab{}.
\newblock \showarticletitle{Do We Learn From Each Other: Understanding the Human-AI Co-Learning Process Embedded in Human-AI Collaboration}.
\newblock \bibinfo{journal}{\emph{Group Decision and Negotiation}} (\bibinfo{year}{2024}), \bibinfo{pages}{1--37}.
\newblock


\bibitem[Muennighoff et~al\mbox{.}(2023)]%
        {muennighoff2023octopack}
\bibfield{author}{\bibinfo{person}{Niklas Muennighoff}, \bibinfo{person}{Qian Liu}, \bibinfo{person}{Armel Zebaze}, \bibinfo{person}{Qinkai Zheng}, \bibinfo{person}{Binyuan Hui}, \bibinfo{person}{Terry~Yue Zhuo}, \bibinfo{person}{Swayam Singh}, \bibinfo{person}{Xiangru Tang}, \bibinfo{person}{Leandro Von~Werra}, {and} \bibinfo{person}{Shayne Longpre}.} \bibinfo{year}{2023}\natexlab{}.
\newblock \showarticletitle{Octopack: Instruction tuning code large language models}.
\newblock \bibinfo{journal}{\emph{arXiv preprint arXiv:2308.07124}} (\bibinfo{year}{2023}).
\newblock


\bibitem[OpenAI(2024a)]%
        {gpto1_reasoning}
\bibfield{author}{\bibinfo{person}{OpenAI}.} \bibinfo{year}{2024}\natexlab{a}.
\newblock \bibinfo{title}{GPT-1o: We've developed a new series of AI models designed to spend more time thinking before they respond. Here is the latest news on o1 research, product and other updates.}
\newblock \bibinfo{howpublished}{\url{https://openai.com/o1/}}.
\newblock
\newblock
\shownote{Accessed: 2025-02-17}.


\bibitem[OpenAI(2024b)]%
        {gpt4o}
\bibfield{author}{\bibinfo{person}{OpenAI}.} \bibinfo{year}{2024}\natexlab{b}.
\newblock \bibinfo{title}{GPT-4o: We’re announcing GPT-4o, our new flagship model that can reason across audio, vision, and text in real time.}
\newblock \bibinfo{howpublished}{\url{https://openai.com/index/hello-gpt-4o/}}.
\newblock
\newblock
\shownote{Accessed: 2025-02-17}.


\bibitem[Park et~al\mbox{.}(2024)]%
        {park2024collaborative}
\bibfield{author}{\bibinfo{person}{Jinkyung~Katie Park}, \bibinfo{person}{Rahul~Dev Ellezhuthil}, \bibinfo{person}{Pamela Wisniewski}, {and} \bibinfo{person}{Vivek Singh}.} \bibinfo{year}{2024}\natexlab{}.
\newblock \showarticletitle{Collaborative Human-AI Risk Annotation: Co-Annotating Online Incivility with CHAIRA}.
\newblock \bibinfo{journal}{\emph{arXiv preprint arXiv:2409.14223}} (\bibinfo{year}{2024}).
\newblock


\bibitem[Pham et~al\mbox{.}(2023)]%
        {pham2023topicgpt}
\bibfield{author}{\bibinfo{person}{Chau~Minh Pham}, \bibinfo{person}{Alexander Hoyle}, \bibinfo{person}{Simeng Sun}, \bibinfo{person}{Philip Resnik}, {and} \bibinfo{person}{Mohit Iyyer}.} \bibinfo{year}{2023}\natexlab{}.
\newblock \showarticletitle{Topicgpt: A prompt-based topic modeling framework}.
\newblock \bibinfo{journal}{\emph{arXiv preprint arXiv:2311.01449}} (\bibinfo{year}{2023}).
\newblock


\bibitem[Weidinger et~al\mbox{.}(2022)]%
        {weidinger2022taxonomy}
\bibfield{author}{\bibinfo{person}{Laura Weidinger}, \bibinfo{person}{Jonathan Uesato}, \bibinfo{person}{Maribeth Rauh}, \bibinfo{person}{Conor Griffin}, \bibinfo{person}{Po-Sen Huang}, \bibinfo{person}{John Mellor}, \bibinfo{person}{Amelia Glaese}, \bibinfo{person}{Myra Cheng}, \bibinfo{person}{Borja Balle}, \bibinfo{person}{Atoosa Kasirzadeh}, {et~al\mbox{.}}} \bibinfo{year}{2022}\natexlab{}.
\newblock \showarticletitle{Taxonomy of risks posed by language models}. In \bibinfo{booktitle}{\emph{Proceedings of the 2022 ACM Conference on Fairness, Accountability, and Transparency}}. \bibinfo{pages}{214--229}.
\newblock


\bibitem[Zhang et~al\mbox{.}(2024)]%
        {zhang2024rethinking}
\bibfield{author}{\bibinfo{person}{Shao Zhang}, \bibinfo{person}{Jianing Yu}, \bibinfo{person}{Xuhai Xu}, \bibinfo{person}{Changchang Yin}, \bibinfo{person}{Yuxuan Lu}, \bibinfo{person}{Bingsheng Yao}, \bibinfo{person}{Melanie Tory}, \bibinfo{person}{Lace~M Padilla}, \bibinfo{person}{Jeffrey Caterino}, \bibinfo{person}{Ping Zhang}, {et~al\mbox{.}}} \bibinfo{year}{2024}\natexlab{}.
\newblock \showarticletitle{Rethinking human-ai collaboration in complex medical decision making: A case study in sepsis diagnosis}. In \bibinfo{booktitle}{\emph{Proceedings of the CHI Conference on Human Factors in Computing Systems}}. \bibinfo{pages}{1--18}.
\newblock


\end{thebibliography}

\end{document}